\documentclass[12pt]{article}
\usepackage{latexsym,epsfig,graphicx,amsmath,amssymb,amscd,undertilde,multirow,chicago,paralist}
\usepackage{array}
\usepackage{float}
\floatstyle{ruled}
\newfloat{algorithm}{tbp}{loa}
\floatname{algorithm}{Algorithm:}
\usepackage{subfig}
\usepackage{color}
\usepackage[british]{babel}
\usepackage{booktabs}
\let\mytoprule\toprule\renewcommand{\toprule}{\mytoprule[0.08em]}
\let\mybottomrule\bottomrule\renewcommand{\bottomrule}{\mybottomrule[0.08em]
}
\let\mymidrule\midrule\renewcommand{\midrule}{\mymidrule[0.07em]
}
\newcommand{\betavec}{{\boldsymbol{\beta}}}

\newcommand{\bvec}{{\boldsymbol{b}}}

\newcommand{\dvec}{{\boldsymbol{d}}}

\newcommand{\Ivec}{{\boldsymbol{I}}}

\newcommand{\wvec}{{\boldsymbol{w}}}

\newcommand{\xivec}{{\boldsymbol{\xi}}}

\newcommand{\N}{{\textrm{N}}}

\newcommand{\alphavec}{{\boldsymbol{\alpha}}}

\newcommand{\xvec}{\boldsymbol{x}}

\newcommand{\lambdavec}{\boldsymbol{\lambda}}

\newcommand{\Rvec}{{\boldsymbol{R}}}
\DeclareMathOperator{\MSE}{MSE}

\begin{document}

\title{Copula-Frailty Models for Recurrent Event Data Based on Monte Carlo EM Algorithm}

\author{Khaled F. Bedair$^{1,2}$, Yili Hong$^3$, and Hussein R. Al-Khalidi$^4$\\
{\small $^{1}$Faculty of Commerce, Tanta University, Tanta, 31521, Egypt}\\
{\small $^{2}$School of Medicine, University of Dundee, Dundee, DD1~9SY, UK}\\
{\small $^{3}$Department of Statistics, Virginia Tech, Blacksburg, VA, 24061, USA}\\
{\small $^{4}$Department of Biostatistics \& Bioinformatics, Duke University, Durham, NC, 27705, USA}
}

%


\date{}
\maketitle

\maketitle

\begin{abstract}
Multi-type recurrent events are often encountered in medical applications when two or more different event types could repeatedly occur over an observation period. For example, patients may experience recurrences of multi-type nonmelanoma skin cancers in a clinical trial for skin cancer prevention. The aims in those applications are to characterize features of the marginal processes, evaluate covariate effects, and quantify both the within-subject recurrence dependence and the dependence among different event types. We use copula-frailty models to analyze correlated recurrent events of different types. Parameter estimation and inference are carried out by using a Monte Carlo expectation-maximization (MCEM) algorithm, which can handle a relatively large (i.e., three or more) number of event types. Performances of the proposed methods are evaluated via extensive simulation studies. The developed methods are used to model the recurrences of skin cancer with different types.

\textbf{Keywords:} Clinical trial; MCEM algorithm; Multi-type recurrences; Multivariate frailty; Skin cancers; Survival models.

\end{abstract}

\newpage

\section{Introduction}
In recurrent event data, the event of interest can occur more than once in a study. Examples of such events include hospitalizations, children's asthma, heart attacks, infections, bleedings, and recurrent tumors. The literature in the analysis of univariate recurrent event data is abundant (e.g., \shortciteNP{prentice1981regression}, \citeNP{andersen1982cox}, and \shortciteNP{wei1989regression}). Sometimes two or more different types of recurrent events may occur throughout the study, and those different types of recurrent events may be correlated to each other. This type of data is referred to as multi-type recurrent event data. For multi-type recurrent event data, it may not be sufficient to perform separate analyses for each type of recurrent event, while ignoring the dependence among event types.

In this paper, we are interested in the modeling and analysis of multi-type recurrent event data, while accounting for the dependence among those event processes. The motivation for this work comes from the clinical trial data collected to study the efficacy of a nutritional supplement of selenium in the prevention of nonmelanoma skin cancers, including basal cell carcinoma (BCC) and squamous cell carcinoma (SCC) (\shortciteNP{duffield2002baseline}, and \shortciteNP{duffield2003selenium}). The analysis aims to describe the features of the marginal processes, examine covariate effects on the risks of different types of recurrent events,  investigate the dependence of within-subject events, and evaluate the correlation among recurrences of different event types. We use copula-frailty models to describe the multi-type recurrent event data, which use the frailty approach to model the dependence of within-subject events of the same type, and use copula to model the dependence of events among different types.

In literature, multivariate frailties/random effects are incorporated into models to accommodate for within-subject event dependence and the dependence among different types of events. A common  assumption is that the  distribution of frailties/random effects belongs to some parametric family, and the normal distribution is used most of the time for modeling random effects (e.g., \shortciteNP{zeng2014multivariate}, \shortciteNP{bedair2016multivariate},  and \shortciteNP{lin2017bayesian}). In addition, the gamma distribution is the most common one for modeling frailties of bivariate survival times (e.g., \citeNP{duchateau_frailty_2008}). Copula models are also used in some cases as alternatives to model bivariate survival data (e.g., \citeNP{shih1995inferences}). \citeN{ChatterjeeRoy2018} used a copula-based approach to estimate the survival functions of two alternating recurrent events.

We choose to use the copula approach based on the following considerations. Copula models involve many multivariate distributions as special cases, which allow the frailty (or random effect) distribution to have more complex features than the symmetric normal density. The multivariate normal distribution can be obtained using the Gaussian copula with normal marginal distributions. \shortciteN{bedair2016multivariate} used a multivariate normal distribution to model multi-type recurrent event data. \citeN{TawiahMcLachlanNg2020} also used multivariate normal distribution to model recurrent events with dependent censoring and cure fraction. However, copula models provide a flexible way to model the dependence structure beside the multivariate normal distribution. Practitioners can have more options in choosing models for describing multi-type recurrent event data.

For the modeling of the baseline intensity function for the frailty model, \shortciteN{rondeau2007joint} used restricted cubic splines for modeling baseline intensity functions. \shortciteN{mazroui2015time} used parametric piecewise and spline baseline intensity functions. \shortciteN{lin2017bayesian} also used parametric piecewise  intensity functions. \citeN{LiGuoKim2020} considered nonparametric Bayesian framework in recurrent event applications. In this paper, the baseline intensity functions are left unspecified. That is, the cumulative baseline function is a step function with jumps only at the observed recurrent event times for each type of event. The unspecified baseline intensity function introduces a layer of difficulty in parameter estimation. However, it can be an attractive feature to practitioners, as often in practice, the shape of the baseline intensity function is unknown.

The parameter estimation of multivariate frailty models can be challenging. A variety of numerical methods have been used to assess the complex integral. Commonly used approaches are the Laplace approximation (e.g., \shortciteNP{cook2010copula}), and the Gaussian quadrature methods (e.g., \shortciteNP{liu2008use}). The Laplace approximation is challenging to implement, especially when the baseline intensity functions are left unspecified because there will be nonparametric terms in the integrands. The quadrature method is hard to scale up to applications with the number of event types being larger than two, which is the bottleneck why most existing multivariate methods only address two types of events. To overcome those difficulties, we use the Monte Carlo expectation-maximization (MCEM) technique for parameter estimation, which can scale up to more than three types of events. \citeN{LeeCook2019} proposed a joint model for multi-type recurrent events using the composite likelihood to approximate the exact likelihood function. We directly handle the full likelihood through the MCEM algorithm.

As discussed above, we aim to develop copula-frailty models with unspecified baseline functions for analyzing multi-type recurrent event data, which have several unique features as compared to existing work in multi-type recurrent event modeling. We use flexible copula models to describe multivariate correlated frailties, which can provide a better fit than existing models. An MCEM algorithm is tailored for estimating parameters, which can handle a relatively large (i.e., three or more) number of event types while existing methods typically handle two types of events. Besides, we provide estimates of fixed effects for each event type, variance components, and the corresponding standard errors for parameter estimators. The dependence of within-subject events and the dependence structure among the marginal processes are obtained.

The remainder of this paper is organized as follows. Section \ref{sec:The-model} introduces some notation for the data and models. Section~\ref{sec:Frailty-models}  introduces copula functions and marginal distributions. Section~\ref{sec:Estimation-Procedure} provides details on the estimation methods using the MCEM algorithms. Extensive simulation studies are conducted to evaluate the performance of the proposed methods in Section~\ref{sec:Simulation-Studies-Co_ch}. Section~\ref{sec:Application-to-Real_Co_ch}  presents an application to the skin cancer data. Section~\ref{sec:Conclusions_Co_ch} contains some conclusions and areas for future research.

\section{Data Setup and Modeling \label{sec:The-model}}

\subsection{Data Setup}
Let $n$ be the number of subjects, and each subject is with a $p$ $\times$ 1 vector of covariates $\xvec_{i}=(x_{i1},\ldots, x_{ip})'$. Subject $i$ is observed over the time interval $[0,\,\tau_{i}]$, where the time is measured from a defined starting point for that subject.  Here $\tau_i$ is the last follow-up time for subject $i$, which is the censoring time. Individuals may experience any of $m$  different  types of recurrent events. Let $t_{ijk}$ be the $k^{th}$ event time of event type $j$ for subject $i$, where $0 < t_{ijk}< \tau_{i}$. For subject $i$, we record $\dvec_{ijk}=(t_{ijk}, 1, \xvec_{i}')', k=1,\ldots, n_{ij}$, where the ``1'' in $\dvec_{ijk}$ is an indicator for an event, and $n_{ij}$ is the number of events of type $j$ from subject $i$. We also include the censoring time information as $\dvec_{ij,(n_{ij}+1)}=(\tau_i, 0, \xvec_{i}')'$, where the ``0'' is an indicator for the censoring time. We then organize our data into matrices. Let $\mathbf{D}_{ij}=(\dvec_{ij1},\ldots,\dvec_{ij, (n_{ij}+1)})'$
be the observed $(n_{ij}+1)\times(p+2)$ data matrix of event
type $j$ for subject $i$, and $\mathbf{D}_{i}=(\mathbf{D}_{i1}',\ldots,\mathbf{D}_{im}')'$ is used to represent the observed data for subject~$i$ over all
$m$ recurrent event types.

\subsection{Multi-type Intensity Model \label{sub:Multi-type-Intensity-Model copula}}
The counting processes are denoted by $N_{ij}(t), i=1,\ldots, n,$ and $j=1,\ldots, m$.
We use $\Delta N_{ij}(t)=N_{ij}(t+\Delta t)-N_{ij}(t)$ to denote the number
of events occurring in the interval $[t,\,t+\Delta t)$.  The event history of subject $i$ by time $t$ is denoted as
$\mathcal{H}_{i}(t)=\{\xvec_{i}, N_{ij}(s),i=1,\ldots,n,$ $\, j=1,\ldots, m, \,0\le s< t\}$. The intensity function of event type $j$ for subject $i$ can be expressed as,
\[
\lambda_{ij}[t|\mathcal{H}_{i}(t)]=\underset{\text{\ensuremath{\Delta}}t
\rightarrow\infty}{\lim}\frac{\mbox{Pr}[\text{\ensuremath{\Delta}}\N_{ij}(t)=1|
\mathcal{H}_{i}(t)]}{\text{\ensuremath{\Delta}}t}\,.
\]
We formulate the multivariate frailty model for intensity
functions $\lambda_{ij}(t_{ijk})$ of recurrent events
as
\begin{equation}\label{eq:model}
\lambda_{ij}(t_{ijk}|w_{ij})=\lambda_{0j}(t_{ijk})\, w_{ij}\,\exp(\xvec_{i}'\betavec_{j}),
\end{equation}
where $\lambda_{0j}(\cdot)$ is a nonparametric baseline intensity
function for the event type $j$. We refer to the model in \eqref{eq:model} as the frailty model. Here, $\betavec_{j}$=$(\beta_{j1},\ldots,\beta_{jp})'$
is a vector for the fixed effects associated with covariates
$\xvec_{i}$ for event type $j$, and $\exp(\xvec_{i}'\betavec_{j})$
models the effect of covariate $\xvec_i$.

Note that in \eqref{eq:model}, we use an intensity-based model to describe recurrent events data, and we incorporate the effects of covariates through the term $\exp(\xvec_{i}'\betavec_{j})$. As the intensity can be viewed as a measure of the risk for an individual to have events, $\exp(\beta_{jl})$ gives the ratio of risks if there is one unit increase in the $l$th covariate given other covariates fixed. In this way,  $\exp(\beta_{jl})$ is usually referred to as the relative risk.

We denote the subject-specific
frailty for the $j^{th}$ event type by $w_{ij}$, and denote the subject-specific
multivariate frailties by $\wvec_{i}=(w_{i1},\ldots, w_{im})'$.
The frailties within subject $i$ are correlated. Note that the regression parameter $\betavec_j$ has conditional interpretation due to the presence of the random variable $w_{ij}$. The baseline cumulative intensity function is defined as
$\Lambda_{0j}(t)=\int_{0}^{t}\lambda_{0j}(s)\, ds$. The cumulative intensity function can be expressed as,
$$\Lambda_{ij}(t)=\Lambda_{0ij}(t)\, w_{ij}\,\exp(\xvec_{i}'\betavec_{j}).$$
The intensity function in \eqref{eq:model} can be rewritten as
\begin{equation}
\lambda_{ij}(t_{ijk}|w_{ij})=\lambda_{0j}(t_{ijk})\,\exp(\xvec_{i}'\betavec_{j}+b_{ij}),\label{eq:log normal model}
\end{equation}
where $b_{ij}=\log(w_{ij})$ is considered to be the random effect for the $j^{th}$ recurrent event type from the $i^{th}$ subject.  We refer to the model in \eqref{eq:log normal model} as the random effects model. Both the frailty and random effects models allow for correlations within subjects.

\section{Copula Frailties/Random Effects Modeling \label{sec:Frailty-models}}
\subsection{Modeling Multivariate Frailties}

Copula functions (e.g., \citeNP{nelsen1999introduction}) are used to model unobserved multivariate frailties and multivariate random effects. That is to model the distribution of frailty vector $\wvec_{i}=(w_{1},\ldots, w_{m})'$ and random effects vector $\bvec_{i}=(b_{1},\ldots, b_{m})'$.  Although our description of the method, henceforward, mainly focuses on the frailty model in \eqref{eq:model}, with some minor modifications, the methods can be applied to the random effects model in \eqref{eq:log normal model}.

In copula modeling, a transformation is made to each marginal variable, $w_{j}$, and
then each transformed marginal variable follows a uniform$(0,1)$ distribution. The dependence
structure is expressed by a multivariate distribution on the transformed
uniform random variables. The copula function $C[F_{1}(w_{1}),\ldots,F_{m}(w_{m})]$
is a joint distribution function such that the joint cumulative distribution function (cdf) can be expressed as
\[
F(w_{1},\ldots,w_{m})=C[F_{1}(w_{1}),\ldots,F_{m}(w_{m})],
\]
with marginal distribution functions $F_{j}(\cdot)$. For simplicity,
we denote $u_{j}=F_{j}(w_{j})$. The copula
density function is defined by
\begin{equation*}
c(u_{1},\ldots, u_{m})=\frac{\partial^{m}}{\partial u_{1},\cdots\,,\partial u_{m}}\, C(u_{1},\ldots, u_{m}).
\end{equation*}
The multivariate probability density function (pdf) of the frailty vector $(w_{1},\ldots, w_{m})'$
is
\begin{equation}
g(w_{1},\ldots, w_{m})=c(u_{1},\ldots, u_{m})\prod_{j=1}^{m}g_{j}(w_{j}).\label{eq:joint pdf}
\end{equation}
The result in~\eqref{eq:joint pdf} shows that it is always possible
to specify a multivariate pdf $g(w_{1},\ldots, w_{m})$
by specifying the marginal pdf $g_{j}(w_{j})$ and
a copula density function $c(\cdot)$.

All information concerning dependence among marginals is contained
in the association parameters. The two most frequently used copula
families are the elliptical and Archimedean copulas, which can be conveniently
used for modeling multivariate frailties.  Some convenient distributions
such as gamma and inverse Gaussian can be used to model marginal frailties, while symmetric distributions such as the normal can be used to model marginal random effects. We briefly introduce the Gaussian and Clayton copulas, and marginal distributions used in copula modeling in the following sections.

\subsection{Gaussian and Clayton Copula}
The multivariate Gaussian copula is given by
\[
C(u_{1},\ldots, u_{m})=\Phi_{m}\big[\Phi^{-1}(u_{1}),\ldots,\Phi^{-1}(u_{m})\big],
\]
where $\Phi_{m}(\cdot)$ and $\Phi(\cdot)$
are the cdf of a multivariate normal distribution $\mbox{MVN}(\mathbf{0},\,\Rvec_{m})$
with a correlation matrix $\Rvec_{m}$, and the standard univariate
normal distribution $\mbox{N}(0,1)$, respectively. The pdf of the normal copula
is given by
\begin{eqnarray*}
	c(u_{1},\ldots, u_{m}) & = & | \Rvec_{m}|^{-\frac{1}{2}}\,\exp\left[-\frac{1}{2}\,\boldsymbol{q}_{i}'(\Rvec_{m}^{-1}-\Ivec_{m})\boldsymbol{q}_{i}\right],
\end{eqnarray*}
where the correlation matrix $\Rvec_{m}$ has $m(m-1)/2$ unique elements to parameterize the dependence of frailties. Here,
$\boldsymbol{u}=(u_{1},\ldots,u_{m})'$, $\boldsymbol{q}=(q_{1},\ldots, q_{m})'$
is a vector of normal scores $q_{j}$= $\Phi^{-1}(u_{j})$, and
$\Ivec_{m}$ is the $m$-dimensional identity matrix. The multivariate pdf for the
frailty can be obtained by $g(w_{1},\ldots, w_{m})$ as in \eqref{eq:joint pdf}.

The one-parameter Clayton copula with the generator function $\psi(u_{j})=u_{j}^{-\alpha}-1$ (i.e., $\psi^{-1}(s)=(1+s)^{-1/\alpha}$) has the following form,
\begin{equation*}
C(u_{1},\ldots, u_{m})=(u_{1}^{-\alpha}+\ldots+u_{m}^{-\alpha}-m+1)^{-\frac{1}{\alpha}},\,\alpha\geq0.
\end{equation*}
Here $\alpha$ is the copula parameter that controls the degree
of dependence. When $\alpha=0$, there is no dependence, and when $\alpha=\infty$
there is perfect dependence. The Kendall's tau can be used as a measurement
for the association by $\tau=\alpha/(\alpha+2)$, which takes values over the interval $[0,\,1]$. The multivariate pdf for $(w_{1},\ldots, w_{m})'$ based on a Clayton copula is
\[
g(w_{1},\ldots, w_{m})=(-\alpha)^{m}\left[\prod_{j=0}^{m-1}\left(-\frac{1}{\alpha}-j\right)\right]
\left[\left(\sum_{j=1}^{m}u_{j}^{-\alpha}-m+1\right)^{-\frac{1}{\alpha}-m}\right]
\left[\prod_{j=1}^{m}u_{j}^{-\alpha-1}g_{j}(w_{j})\right].
\]
\subsection{Marginal Distributions}
For the gamma frailty model, the frailty $w_{j}$ is distributed with
gamma$(1/\alpha_{j},\,\alpha_{j})$ with mean one and variance
$\alpha_{j}$, and the pdf is
\[
g_{j}(w_{j})=\frac{w_{j}^{(1/\alpha_{j}-1)}\,\exp(-w_{j}/\alpha_{j})}{\Gamma(1/\alpha_{j})\alpha_{j}^{1/\alpha_{j}}}.
\]
For the lognormal frailty model, $w_{j}=\exp(b_{j})$, and $b_{j}\thicksim\mbox{N}(0,\,\alpha_{j})$ is normally
distributed with mean zero and variance $\alpha_{j}$. That is, $w_{j}$ has the lognormal distribution. The mean and variance of the frailty are $\mbox{E}(w_{j})=\exp(\alpha_{j}/2)$ and $\mbox{Var}(w_{j})=\exp(\alpha_{j})[\exp(\alpha_{j})-1]$, respectively.

\section{Statistical Inference\label{sec:Estimation-Procedure}}

\subsection{The Log-likelihood Function}
Let $\betavec=(\betavec_{1}',\ldots,\betavec_{m}')'$
and $\alphavec=(\alpha_{1},\ldots,\alpha_{m},\alphavec_c')'$ be the parameter vectors.
Here $\alphavec_c$ denotes the parameter(s) in the copula function. For the Clayton copula, $\alphavec_c=\alpha$. For the Gaussian copula, $\alphavec_c$ denotes the parameters in the correlation matrix $\Rvec_{m}$. The cumulative baseline function is specified as a step function with jumps only at the observed recurrent event times for each of the $m$ types of events. The ordered distinct event times are denoted by $t_{j(1)},\ldots,t_{j(k_{j})}$, where $k_{j}$ is the number of distinct event times from type $j$ for all subjects. The corresponding baseline intensity functions can be represented as $\lambdavec_{0j}=\left\{\lambda_{0j}[t_{j(1)}],\ldots,\allowbreak\lambda_{0j}[t_{j(k_{j})}]\right\}'$. We then denote the unspecified baseline intensity functions as $\lambdavec_{0}=(\lambdavec_{01}',\ldots,\lambdavec_{0m}')'$. The vector of unknown parameters to be estimated in the model is $\xivec=(\betavec',\,\lambdavec_{0}',\,\alphavec')'.$

We denote the data for subject $i$ by $\mathbf{D}_{i}$ and frailty terms by $\wvec_{i}$. The multivariate pdf of the frailty terms $\wvec_{i}$ is $g(\wvec_i|\alphavec)=g(w_{i1},\ldots\,,w_{im})$. Given data $\mathbf{D}_i$ and $\wvec_{i}$, one can write down the $i^{th}$ subject's
contribution to the likelihood function as,
\begin{eqnarray}
L_{i}(\xivec,\wvec_{i}) & = & \prod_{j=1}^{m}\left[\prod_{k=1}^{n_{ij}}\lambda_{j}(t_{ijk})\right]\exp\big[-\Lambda_{j}(\tau_{i})\big]g(\wvec_{i}|\alphavec).
\label{eq:subject log}
\end{eqnarray}
By substituting $\lambda_{j}(t_{ijk})$ and $\Lambda_{j}(\tau_{i})$ as in \eqref{eq:model} and taking the logarithm over all subjects, we can obtain the complete log-likelihood as $\mathcal{L}(\xivec,\wvec)=\sum_{i=1}^{n}\log[L_{i}(\xivec,\wvec_{i})]$.
In particular,
\begin{align}\nonumber
\mathcal{L}(\xivec,\wvec)=&\sum_{i=1}^n\sum_{j=1}^{m}\Bigg( \sum_{k=1}^{n_{ij}}\Big\{\log\,\big[\lambda_{0j}(t_{ijk})\big]+\log(w_{ij})+\xvec_{i}'\betavec_{j}\Big\}-\Lambda_{0j}(\tau_{i})w_{ij}\exp(\xvec_{i}'\betavec_{j})\Bigg) \\
&+\sum_{i=1}^n\log\big[g(\wvec_{i}|\alphavec)\big],\label{eq:log likelihood}
\end{align}
where $\wvec=(\wvec_{1}',\ldots,\wvec_n')'$
is the frailty vector all over $\wvec_{i}$. The marginal likelihood of the observed data over all subjects
is
\begin{equation}
L(\xivec)=\prod_{i=1}^n\int_{\wvec_{i}}L_{i}(\xivec,\wvec_{i})d\wvec_{i}.\label{eq:marginal}
\end{equation}
In most cases, this integration does not have a closed-form expression. The
likelihood function in \eqref{eq:marginal} has two challenges in obtaining the inference of $\xivec$. First, it depends on the high dimensional nonparametric baseline intensity function. Second, it is usually a multi-dimensional integration. We use the MCEM technique to overcome those two difficulties.

\subsection{Monte Carlo EM Algorithm}
In the MCEM algorithm, the expectation in the E-step is computed using Monte Carlo simulations. The MCEM algorithm includes two steps: the computing of conditional expectations for the log of the complete
likelihood (E-step) and the maximization of the conditional expectations with respect
to all parameters (M-step). With initial values, the algorithm does iterations between the two steps. As indicated in \citeN{booth1999maximizing}, the algorithm converges to a stationary point under certain regularity conditions.

\subsubsection{E-step}
In the E-step, the pdf of $\wvec_{i}$ conditional
on observed data is
\begin{equation}
g_{\wvec_{i}|\mathbf{D}_{i}}(\wvec_{i}|\xivec)=
g(\wvec_{i}|\boldsymbol{\xivec})=\frac{f(\mathbf{D}_{i},\wvec_{i})}
{\intop_{\wvec_{i}}f(\mathbf{D}_{i},\wvec_{i})d\wvec_{i}}
=\frac{L_{i}(\xivec,\wvec_{i})}{L_{i}(\xivec)},\label{eq:marginal frai}
\end{equation}
where $f(\mathbf{D}_{i},\wvec_{i})$ is the joint density
of the data and frailty, $L_{i}(\xivec,\wvec_{i})$
is defined in \eqref{eq:subject log}, and $L_{i}(\xivec)$
is the marginal likelihood for the $i^{th}$ subject. In the E-step,
since there is no closed-form for the density $g_{\wvec_{i}|\mathbf{D}_{i}}(\wvec_{i}|\xivec)$,
the Metropolis-Hastings algorithm is used to generate random
samples of $\wvec_{i}$ with the conditional distribution
in \eqref{eq:marginal frai}. A description to the Metropolis-Hastings
algorithm is given in Appendix~\ref{sec:B.1Metropolis-Hasting-Algorithm}.
For each subject $i$, we generate random
samples (after the burn-in and thinning) $\wvec_{i}^{(q)}$, $q=1,\ldots, n_s$. Then the
expectations of functions of $\wvec_{i}$ conditional on the
observed data are computed by averaging of the $n_s$ samples. That is,
\[
\mbox{E}(w_{ij})=\frac{1}{n_s}\sum_{q=1}^{n_s}w_{ij}^{(q)},\,\quad \mbox{and}\quad\,\mbox{E}\big[\log (w_{ij})\big]=\frac{1}{n_s}\sum_{q=1}^{n_s}\big[\log(w_{ij}^{(q)})\big].
\]

\subsubsection{M-step}
The EM algorithm requires $Q(\xivec)$, which
is the expectation of the log-likelihood in \eqref{eq:log likelihood} conditional on all the data and
current parameter estimates. In particular, $Q(\xivec)$
can be expressed as
\[
Q(\xivec)=Q_{1}(\betavec, \lambdavec_{0})+Q_{2}(\alphavec),
\]
where
\begin{align}
Q_{1}(\betavec, \lambdavec_{0})=&\sum_{i=1}^n\sum_{j=1}^{m} \sum_{k=1}^{n_{ij}}\log[\lambda_{0j}(t_{ijk})]+\mbox{E}[\log(w_{ij})]+\xvec_{i}'\betavec_{j}\\\nonumber
&-\sum_{i=1}^n\sum_{j=1}^{m}\Lambda_{0j}(\tau_{i})\mbox{E}(w_{ij})\exp(\xvec_{i}'\betavec_{j}),
\end{align}
and $Q_{2}(\alphavec)=\sum_{i=1}^n\mbox{E}\{\log[g(\wvec_{i}|\alphavec)]\}.$

The regression parameters are updated by  maximizing the expected partial likelihood.
In particular,
\begin{equation} \mbox{E}[\mathcal{L}_{partial}(\betavec)]=\sum_{j=1}^{m}\sum_{l=1}^{k_{j}}\left\{\xvec_{i}'\betavec_{j}+\mbox{E}[\log( w_{ij})]-\log\left[\underset{i\in R(t_{j(l)})}{\sum}\mbox{E}(w_{ij})\exp(\xvec_{i}'\betavec_{j})\right]\right\},
\end{equation}
where $R(t_{j(l)})$ is the at risk group of event type $j$ at time $t_{j(l)}$. Note that the $\mbox{E}[\mathcal{L}_{partial}(\betavec)]$ function is separable for $\betavec_j$'s. Thus the maximization of $\mbox{E}[\mathcal{L}_{partial}(\betavec)]$ can be done separable for each $j$, which can reduce the complexity of the optimization problem and allows for a relatively large number of types of events. The cumulative intensity functions $\Lambda_{0j}(\cdot)$ for the
recurrent events can be updated by
\begin{equation}
\widehat{\Lambda}_{0j}(t)=\sum_{t_{j(l)}\leq t}\frac{N_{j}(t_{j(l)})}{\underset{i\in R(t_{j(l)})}{\sum}\mbox{E}(w_{ij})\exp(\xvec_{i}'\betavec_{j})},
\end{equation}
where $N_{j}(t_{j(l)})$ is the total number of
events of type $j$ at time $t_{j(l)}$.

The estimation of copula parameter $\alphavec$
can be obtained by maximizing $Q_{2}(\alphavec)$,
where
\begin{align*}
Q_{2}(\alphavec)= & \mbox{E}\left\{ \log[g(w_{i1},\ldots, w_{im})]\right\} =\mbox{E}\left\{ \log\left[c(u_{i1},\ldots, u_{im})\prod_{j=1}^{m}g_{j}(w_{ij})\right]\right\} \\
	 = & \mbox{E}\{\log[c(u_{i1},\ldots, u_{im})]\}+\sum_{j=1}^{m}\mbox{E}\{\log[g_{j}(w_{ij})]\}=Q_{3}(\alphavec_c)+\sum_{j=1}^{m}Q_{4}(\alpha_{j}).
\end{align*}
Here, $Q_{3}(\alphavec_c)$ and $Q_{4}(\alpha_{j})$ denote the expectation of the log-likelihood of the copula and the marginal distributions, respectively.
The estimation of $\alphavec$ can be achieved by maximizing
$Q_{2}(\alphavec)$. In particular, the forms
of the expected log-likelihood for Gaussian and Clayton copula with gamma marginals and their score equations are derived in Appendix B. For the estimation of the parameters in the copula function and marginal distributions, we apply a two-stage estimation method as commonly used in literature for copula models (e.g., \citeNP{JoeXu1996}, and \citeNP{Joe2005}), which estimates the marginal distribution parameters $\alpha_j$'s in the first step by maximizing $Q_4(\alpha_j)$, and then estimates the copula association parameter $\alphavec_c$ by maximizing $Q_3(\alphavec_c)$ given $\widehat{\alpha}_j$'s. As a result, the first and second derivatives with respect to $\alpha_j$ have been taken to the related likelihood $Q_4(\alpha_j)$, which is independent from $Q_3(\alphavec_c)$. One advantage of the two-stage method is that it reduces the complexity of the optimization problem, allowing to estimate parameters for models with more than three types of events. As a summary, the MCEM algorithm is outlined in \textbf{Algorithm~\ref{alg:Summary-of-the MCMC}}.

\begin{algorithm}
	\caption{\label{alg:Summary-of-the MCMC} Monte Carlo EM algorithm for copula-frailty model.}
	\begin{enumerate}
		\item Initialize $\widehat\xivec^{(0)}$. At iteration $(s+1)$,
		\item E-step:
		\begin{enumerate}
			\item Generate $\wvec=(\wvec_{1}',\ldots,\wvec_n')'\thicksim g_{\wvec_{i}|\mathbf{D}_{i}}(\wvec_{i}|\widehat\xivec^{(s)})$
			via a Markov chain Monte Carlo (MCMC) algorithm.
			\item Compute the required conditional expectations $\mbox{E}\left\{ g_{\wvec_{i}|\mathbf{D}_{i}}[\wvec_{i}|\widehat\xivec^{(s)}]\right\} $
			of the frailty terms.
		\end{enumerate}
		\item M-step: Maximize the expected complete log-likelihood $Q(\xivec)$
		to obtain $\widehat{\xivec}^{(s+1)}$.
		\item Repeat Steps $2$ and $3$ until the convergence is declared.
	\end{enumerate}
\end{algorithm}

The algorithm is stopped and the convergence is declared at the $(s+1)$th step if
\[
\underset{d}{\max}\Bigg(\Big|\frac{\widehat{\xi}_{d}^{(s+1)}-\widehat{\xi}_{d}^{(s)}}{\widehat{\xi}_{d}^{(s)}-\delta_{1}}\Big|\Bigg)<\delta_{2},
\]
where the maximum is taken over all the coordinates of parameter  vector $\xivec$, $\xi_d$ is the $d^{\,th}$ coordinate of $\xivec$, and $\delta_{1},\,\delta_{2}$ are pre-specified small values (e.g.,$\delta_{1}=0.01,\,\delta_{2}=0.003$) as suggested in \citeN{booth1999maximizing}. In practice, such criteria can be at the risk of terminating too early, as it may be obtained only because of Monte Carlo error in the updates. To avoid this implication, the algorithm is  terminated after  such criterion is achieved for three consecutive iterations. Graphical tools such as the trace plots can be used to check the convergence.

For the MCEM algorithm, the Louis formula (\citeNP{louis1982finding}) is needed to provide the information matrix, $I(\widehat{\xivec})$, which is given as follows,
\begin{equation}
I(\widehat{\xivec})=\mbox{E}\left(-\frac{\partial^{2}\mathcal{L}}{\partial\xivec\partial\xivec'}\bigg|\wvec,\widehat{\xivec}\right)
-\mbox{E}\left(\frac{\partial\mathcal{L}}{\partial\xivec}\frac{\partial\mathcal{L}}{\partial\xivec'}\bigg|\wvec,\widehat{\xivec}\right).
\end{equation}
The expectations are computed by averaging over the terms involving samples from MCMC. Theorem~3 of \citeN{parner1998asymptotic} showed that the variance
of $\widehat{\betavec}$, $\widehat{\alphavec}$, and $\widehat{\lambdavec}_{0j}$ can be consistently estimated by using the discrete information matrix, which is the negative of the Hessian matrix. The negative Hessian matrix is obtained by taking second derivatives with respect to parameters ($\betavec$ and $\alphavec$) and the jumps $\lambda_{0j}(t_{j(l)})$.

\section{Simulation Studies\label{sec:Simulation-Studies-Co_ch}}
In this section, we use extensive simulations to study the performance of the proposed methods.

\subsection{Simulation Setting and Data Generation}
Here we discuss our simulation settings. Table~\ref{tab:model.label} gives the list of copula functions and marginal distributions under consideration. Note that when the Gaussian (normal) distribution is used as the marginal distribution, it is more convenient to specify it in the random effects model. As shown in the table, we consider four distribution settings, namely, the Cg, CG, Gg, and GG models. We consider three types of events (i.e., $m=3$) in the simulation, which is general enough to show the scalability of our methods. Our methods can be easily applied to cases where the number of event types is more than three, because of the separability of $\betavec_j$'s in the partial likelihood function and the two-stage estimation method for the copula parameters.

We set the Clayton copula parameter to be $0.1,$ $1.333$, and $8$, which is equivalent to the Kendall's
tau as $0.05,\,0.4$, and $0.8$, respectively. For the Gaussian copula, there are three parameters we can set when $m=3$. To simplify the setting, we set the correlation parameters to be identical. That is, we set $\rho=\rho_{12}=\rho_{13}=\rho_{23}$. In the simulation, the value of the Gaussian copula parameter is set as $\rho=0, 0.4$, and $0.8$ to achieve different degrees of correlation. Although we set the true values of the correlations to be identical, this is not a constraint on the parameter estimates. That is, we still estimate the three correlations (i.e. $\rho_{12}, \rho_{13}$, and $\rho_{23}$) in the estimation procedure.

For the marginal distributions, we consider two choices, the gamma distribution and the Gaussian (normal) distribution. For the gamma distribution, it is typically used in the frailty model, that is $w_{ij}\sim\Gamma(1/\alpha_{j},\,\alpha_{j}).$ For the Gaussian distribution, it is typically used in the random effects model, that is $b_{ij}\sim\mbox{N}(0,\,\alpha_{j})$. We set $\alpha_1=\alpha_2=\alpha_3=1$ in the simulation.

For simplicity, the true baseline hazard functions are set as $\lambda_{0j}(t)=1$, The baseline hazard functions, however, are estimated nonparametrically. We consider the treatment variable as the covariate (i.e., we set 0 as the placebo and 1 as the treatment) for simplicity for not losing insights. Our estimation method is general and works for the situation that $\xvec_i$ is a vector. The effects of covariate are $\beta_{1}=1$, $\beta_{2}=0.8$, and $\beta_{3}=0.4$ for different types of events. We carry out simulation studies for sample size $n=200$ and $400$ subjects. Between 29\% and 33\% of the simulated subjects were censored without developing events of any types. For each setting, 1000 simulated samples
were generated to calculate the results. All computing for the simulation studies was conducted in R.

\begin{table}
\caption{List of copula functions and marginal distributions under consideration. Note that when the Gaussian (normal) distribution is used as the marginal distribution, it is more convenient to specify it in the random effects model.}\label{tab:model.label}
\centering
\begin{tabular}{ccccc}\toprule
Model Label &\quad& Copula Function &\quad& Marginal Distribution\\\midrule
Cg && Clayton  && gamma frailty      \\\hline
CG && Clayton  && Gaussian random effect  \\\midrule
Gg && Gaussian && gamma frailty     \\\hline
GG && Gaussian && Gaussian random effect  \\\bottomrule
\end{tabular}
\end{table}

In summary, we use the following steps to generate the multi-type recurrent event data.
\begin{enumerate}
	\item For subject $i$, where $i=1,\ldots, n$, we generate the frailty $\wvec_i$ (or the random effect $\bvec_i$) according to the chosen copula model.
\item The subjects are randomly assigned to the treatment group with equal probability.
\item The maximum follow up time was set at $C=1,\, i=1,\cdots,n$.
\item Set $\tau_{i}$ as the censored time for subject $i^{th}$, where $\tau_{i}=\min\left\{C_{i}^{\ast},C\right\}$
	and the random censoring time $C_{i}^{\ast}$ is assumed to be exponentially
	distributed with rate $0.5$.
\item For each event type $j$, generate gap times $z_{ijl}$
	from the exponential distribution with $\lambda_{0j}(t)=1$. The
	rate parameter is set as $[\lambda_{0j}(t)w_{ij}\exp(\xvec_{i}'\betavec_{j})]^{-1}$ for the model in (\ref{eq:model}), and is set as  $[\lambda_{0j}(t)\exp(\xvec_{i}'\betavec_{j}+b_{ij})]^{-1}$
	for  the model in (\ref{eq:log normal model}).
\item Set $y_{ijk}=\sum_{l=1}^{n_{ij}}z_{ijl}$
	and let the event time be $t_{ijk}=\min\{\tau_{i},\,$ $y_{ijk}\}$,
	with the first start time set to be zero.
\end{enumerate}
\subsection{Simulation Results}
In the simulation results, the bias is measured as the mean of the parameter estimates
(based on 1000 repeats) minus the true value, and the variance is the sampling
variance of the parameter estimates. The $\MSE$ represents the empirical
mean squared errors of the corresponding parameter estimates over all 1000
repeats. We also compute the empirical coverage probabilities (CP)
of the corresponding 95\% confidence intervals. Tables~\ref{tab:Empirical-standard-errors-Claton_Gamma}, \ref{tab:Empirical-standard-errors-Gaussian_Clayton}, \ref{tab:Empirical-standard-errors-Gaussian_Gamma}, and \ref{tab:Empirical-standard-errors-Gaussian_Gaussian} present the simulation results for the Cg, CG, Gg, and GG models, respectively.

\begin{table}
	\centering
	\caption{\label{tab:Empirical-standard-errors-Claton_Gamma}Empirical results
		from simulation studies examining the properties of the estimators
		of treatment effects, variances, and copula parameter for the Cg model (i.e., Clayton copula with  gamma
		marginal distribution.)}
	{\small
		\begin{tabular}{clrrrrrrrrrr}
			\toprule
			Param. & Value  & \multicolumn{2}{c}{Mean} & \multicolumn{2}{c}{Bias} & \multicolumn{2}{c}{Var} & \multicolumn{2}{c}{$\MSE$} & \multicolumn{2}{c}{CP}\tabularnewline
			\midrule
			\multicolumn{2}{c}{\# of subjects} & 200 & 400 & 200 & 400 & 200 & 400 & 200 & 400 & 200 & 400\tabularnewline
			\midrule
			\multicolumn{12}{c}{Setting I}\tabularnewline
			$\beta_{1}$ & 1 & 0.959 & 1.047 & $-$0.041 & 0.047 &  0.120  &  0.003 & 0.122  &  0.005 & 0.964& 0.951\tabularnewline
			$\beta_{2}$ & 0.8 &0.869&0.795 &0.069 &$-$0.005 &0.116 &0.074 &0.121 &0.074&0.939&0.943\tabularnewline
			$\beta_{3}$ & 0.4 & 0.449 &0.449 &0.049 &0.049  &0.099 &0.057 &0.101 &0.060 &0.941& 0.945\tabularnewline
			$\alpha_{1}$ & 1 &0.951 &0.963& $-$0.049 & $-$0.037& 0.090& 0.060& 0.092& 0.061& 0.933& 0.953\tabularnewline
			$\alpha_{2}$ & 1 &0.978&0.977 &$-$0.022 &$-$0.023 &0.082 &0.039 &0.083 &0.039& 0.962& 0.941\tabularnewline
			$\alpha_{3}$ & 1 &0.975 &0.960 &$-$0.025 &$-$0.040 &0.096 &0.020 &0.096 &0.021& 0.941& 0.942\tabularnewline
			$\alpha$ & 0.1 & 0.217 &0.106 &0.117 &0.006 &0.107  &0.026 &0.121 &0.026& 0.920& 0.934\tabularnewline
			\midrule
			\multicolumn{12}{c}{Setting II}\tabularnewline
			$\beta_{1}$ & 1 & 1.002 &1.011  & 0.002 & 0.011 & 0.126& 0.092 & 0.126 &0.092& 0.933& 0.949\tabularnewline
			$\beta_{2}$ & 0.8 & 0.812 &0.842 &0.012 &0.042 &0.125 &0.038 &0.125 &0.040 &0.937& 0.946\tabularnewline
			$\beta_{3}$ & 0.4 & 0.426 &0.433 &0.026 &0.033 &0.058 &0.028 &0.059 &0.029&0.943 &0.945\tabularnewline
			$\alpha_{1}$ & 1 & 0.998 &0.980  &$-$0.002 &$-$0.020 &0.088  &0.042  &0.088  &0.042& 0.932& 0.948\tabularnewline
			$\alpha_{2}$ & 1 & 1.006 & 0.976&0.006 &$-$0.024 &0.118 &0.087&0.118 &0.088& 0.943& 0.951\tabularnewline
			$\alpha_{3}$ & 1 & 1.011 &1.045 &0.011  &0.045 &0.037 &0.032 &0.037 &0.034& 0.945& 0.949\tabularnewline
			$\alpha$ & 1.333 & 1.370 &1.406  &0.037  &0.073 &0.026  &0.023  &0.027  &0.029& 0.945& 0.946\tabularnewline
			
			\midrule
			\multicolumn{12}{c}{Setting III}\tabularnewline
			$\beta_{1}$ & 1 &1.057&0.970  & 0.057& $-$0.030 & 0.102&  0.052 & 0.105& 0.053& 0.939& 0.956\tabularnewline
			$\beta_{2}$ & 0.8 &0.830& 0.777 &0.030& $-$0.023 &  0.016& 0.009 & 0.017 & 0.010& 0.941& 0.947\tabularnewline
			$\beta_{3}$ & 0.4 &0.461 &0.439 &0.061 &0.039 &0.063 &0.032 &0.067 &0.033& 0.946& 0.949\tabularnewline
			$\alpha_{1}$ & 1 & 0.972 &1.047  &$-$0.028  &0.047 &0.136 &0.093 &0.137 &0.095&0.947& 0.954\tabularnewline
			$\alpha_{2}$ & 1 &  1.039 &0.970 &0.039 &$-$0.030  &0.093 &0.062 &0.094 &0.063 &0.956& 0.956\tabularnewline
			$\alpha_{3}$ & 1 & 0.952 &1.024 &$-$0.048 &0.024 &0.092 &0.025 &0.094 &0.026& 0.941& 0.951\tabularnewline
			$\alpha$ & 8 & 8.736 &7.719 &0.736 &$-$0.281 &0.078 &0.026 &0.620 &0.105 &0.936& 0.943\tabularnewline
			\bottomrule
		\end{tabular}
	}
\end{table}

We focus on discussing the results in Table~\ref{tab:Empirical-standard-errors-Claton_Gamma} first. From the results in Table~\ref{tab:Empirical-standard-errors-Claton_Gamma}, we observe the following:
\begin{enumerate}
\item As expected, the $\MSE$ of all parameter estimates is generally decreasing as the number of subjects increases. A similar pattern is observed for the bias and the variance as we increase the number of subjects.
\item Higher correlations cases of frailties tend to have a higher $\MSE$ for the variance components and the copula parameter. However, the $\MSE$ of the regression coefficients is not affected by the increase of the association parameter.
\item The performance of the CP is around the 0.95 nominal level, and it is getting closer to 0.95 when the number of subjects increases. A slight bias is noted for the estimate of copula parameter with improvements for cases with a larger number of subjects.
\end{enumerate}

For the additional simulation results for the CG, Gg, and GG models, similar conclusions can be reached for those three cases with negligible bias for the estimator of the copula parameter and the CP of the intervals. In summary, the simulation results show that our procedure works well in terms of parameter estimation and statistical inference.

\begin{table}
	\caption{\label{tab:Empirical-standard-errors-Gaussian_Clayton}Empirical results
		from simulation studies examining the properties of the estimators
		of treatment effects, variances, and copula parameter for the CG model (i.e., Clayton copula with Gaussian
		marginal distributions). }
	\centering{}%
	{\small
	\begin{tabular}{clrrrrrrrrrr}
	\toprule
		Param. & Value  & \multicolumn{2}{c}{Mean} & \multicolumn{2}{c}{Bias} & \multicolumn{2}{c}{Var} & \multicolumn{2}{c}{$\MSE$} & \multicolumn{2}{c}{CP}\tabularnewline
		\midrule
		\multicolumn{2}{c}{\# of subjects} & 200 & 400 & 200 & 400 & 200 & 400 & 200 & 400 & 200 & 400\tabularnewline
		\midrule
		\multicolumn{12}{c}{Setting I}\tabularnewline
		$\beta_{1}$ & 1 &  0.963& 1.048 & $-$0.037 & 0.048 & 0.105 & 0.050 & 0.106 & 0.052 & 0.939 & 0.946\tabularnewline
		$\beta_{2}$ & 0.8 &0.791 & 0.845 & $-$0.009 & 0.045 & 0.118 & 0.070 & 0.118 & 0.072 & 0.943 & 0.949\tabularnewline
		$\beta_{3}$ & 0.4 & 0.424 & 0.424 & 0.024 & 0.024 & 0.094 & 0.020 & 0.094 & 0.021 & 0.968 & 0.953\tabularnewline
		$\alpha_{1}$ & 1 & 0.961 & 0.977 & $-$0.039 & $-$0.023 & 0.093 & 0.087 & 0.095 & 0.087 & 0.969 & 0.948\tabularnewline
		$\alpha_{2}$ & 1 &1.053 & 0.967 & 0.053 & $-$0.033 & 0.090 & 0.008 & 0.092 & 0.009 & 0.949 & 0.950\tabularnewline
		$\alpha_{3}$ & 1 & 0.961 & 0.961 & $-$0.039 & $-$0.039 & 0.082 & 0.064 & 0.083 & 0.065 & 0.930 & 0.946\tabularnewline
		$\alpha$ & 0.1 & 0.131 & 0.113 & 0.031 & 0.013 & 0.097 & 0.038 & 0.098 & 0.038 & 0.946 & 0.944 \tabularnewline
		\midrule
		\multicolumn{12}{c}{Setting II}\tabularnewline
		$\beta_{1}$ & 1 & 1.068 & 0.981 & 0.068 & $-$0.019 & 0.132 & 0.031 & 0.136 & 0.032 & 0.940 & 0.946\tabularnewline
		$\beta_{2}$ & 0.8 & 0.790 & 0.827 & $-$0.010 & 0.027 & 0.127 & 0.035 & 0.127 & 0.035 & 0.950 & 0.953\tabularnewline
		$\beta_{3}$ & 0.4 & 0.432 & 0.429 & 0.032 & 0.029 & 0.082 & 0.038 & 0.083 & 0.038 & 0.950 & 0.951\tabularnewline
		$\alpha_{1}$ & 1& 0.931 & 0.971 & $-$0.069 & $-$0.039 & 0.082 & 0.028 & 0.087 & 0.030 & 0.935 & 0.957\tabularnewline
		$\alpha_{2}$ & 1 & 0.984 & 0.952 & $-$0.016 & $-$0.048 & 0.093 & 0.079 & 0.094 & 0.082 & 0.955 & 0.965\tabularnewline
		$\alpha_{3}$ & 1 &0.957 & 0.970 & $-$0.043 & $-$0.030 & 0.059 & 0.014 & 0.061 & 0.015 & 0.949 & 0.951 \tabularnewline
		$\alpha$ & 1.333 & 1.557 & 1.484 & 0.224 & 0.151 & 0.049 & 0.027 & 0.099 & 0.050 & 0.941 & 0.945\tabularnewline
		
		\midrule
		\multicolumn{12}{c}{Setting III}\tabularnewline
		$\beta_{1}$ & 1 & 1.059 & 1.018 & 0.059 & 0.018 & 0.122 & 0.051 & 0.125 & 0.051 & 0.940 & 0.951\tabularnewline
		$\beta_{2}$ & 0.8 & 0.814 & 0.823 & 0.014 & 0.023 & 0.123 & 0.044 & 0.123 & 0.045 & 0.945 & 0.959\tabularnewline
		$\beta_{3}$ & 0.4& 0.420 & 0.430 & 0.020 & 0.030 & 0.107 & 0.017 & 0.108 & 0.018 & 0.944 & 0.954\tabularnewline
		$\alpha_{1}$ & 1  &1.012 & 1.048 & 0.012 & 0.048 & 0.134 & 0.102 & 0.134 & 0.104 & 0.931 & 0.960\tabularnewline
		$\alpha_{2}$ & 1 & 0.987 &0.978  & $-$0.013 & $-$0.022 & 0.129 & 0.057 & 0.129 & 0.058 & 0.933 & 0.953\tabularnewline
		$\alpha_{3}$ & 1 & 0.939 & 1.046 & $-$0.061 & 0.046 & 0.114 & 0.095 & 0.118 & 0.097 & 0.954 & 0.955\tabularnewline
		$\alpha$ & 8 & 7.004 & 8.278 & $-$0.996 & 0.278 & 0.101 & 0.124 & 1.093 & 0.201 & 0.938 & 0.947\tabularnewline
		\bottomrule
	\end{tabular}}
\end{table}

\begin{table}
	\caption{\label{tab:Empirical-standard-errors-Gaussian_Gamma}Empirical results
		from simulation studies examining the properties of the estimators
		of treatment effects, variances, and copula parameter for the Gg model (i.e., Gaussian copula with gamma
		marginal distributions). }

	\noindent \raggedright{}%
	{\small
	\begin{tabular}{clrrrrrrrrrr}
	\toprule
		Param. & Value  & \multicolumn{2}{c}{Mean} & \multicolumn{2}{c}{Bias} & \multicolumn{2}{c}{Var} & \multicolumn{2}{c}{$\MSE$} & \multicolumn{2}{c}{CP}\tabularnewline
		\midrule
		\multicolumn{2}{c}{\# of subjects} & 200 & 400 & 200 & 400 & 200 & 400 & 200 & 400 & 200 & 400\tabularnewline
		\midrule
		\multicolumn{12}{c}{Setting I}\tabularnewline
		$\beta_{1}$ & 1& 1.033 & 0.965 & 0.033 & $-$0.035 & 0.091 & 0.031 & 0.092 & 0.032 & 0.956 & 0.959\tabularnewline
		$\beta_{2}$ & 0.8 & 0.809 & 0.847 & 0.009 & 0.047 & 0.124 & 0.017 & 0.124 & 0.019 & 0.947 & 0.951 \tabularnewline
		$\beta_{3}$ & 0.4  & 0.464 & 0.363 & 0.064 & $-$0.037 & 0.087 & 0.015 & 0.092 & 0.016 & 0.955 & 0.961\tabularnewline
		$\alpha_{1}$ & 1& 0.937 & 0.996 & $-$0.063 & $-$0.004 & 0.062 & 0.065 & 0.066 & 0.065 & 0.946 & 0.954 \tabularnewline
		$\alpha_{2}$ & 1& 0.978 & 0.960 & $-$0.022 & $-$0.040 & 0.033 & 0.019 & 0.033 & 0.020 & 0.950 & 0.949\tabularnewline
		$\alpha_{3}$ & 1& 1.026 & 1.028 & 0.026 & 0.028 & 0.095 & 0.021 & 0.096 & 0.022 & 0.945 & 0.955\tabularnewline
		$\rho_{12}$ & 0 & 0.054 & 0.037 & 0.054 & 0.037 & 0.062 & 0.059 & 0.065 & 0.060 & 0.941 & 0.944\tabularnewline
		$\rho_{13}$ & 0& 0.004 & 0.009 & 0.004 & 0.009 & 0.082 & 0.059 & 0.083 & 0.059 & 0.944 & 0.952\tabularnewline
		$\rho_{23}$ & 0 & 0.056 & 0.020 & 0.056 & 0.020 & 0.101 & 0.054 & 0.104 & 0.055 & 0.946 & 0.951\tabularnewline
		\midrule
		\multicolumn{12}{c}{Setting II}\tabularnewline
		$\beta_{1}$ & 1 & 1.066 & 1.043 & 0.066 & 0.043 & 0.082 & 0.006 & 0.087 & 0.008 & 0.950 & 0.956\tabularnewline
		$\beta_{2}$ & 0.8& 0.822 & 0.849 & 0.022 & 0.049 & 0.106 & 0.084 & 0.106 & 0.086 & 0.945 & 0.956\tabularnewline
		$\beta_{3}$ & 0.4 & 0.413 & 0.433 & 0.013 & 0.033 & 0.134 & 0.094 & 0.135 & 0.095 & 0.950 & 0.966 \tabularnewline
		$\alpha_{1}$ & 1& 1.068 & 1.022 & 0.068 & 0.022 & 0.060 & 0.050 & 0.064 & 0.050 & 0.949 & 0.958\tabularnewline
		$\alpha_{2}$ & 1 & 1.062 & 1.019 & 0.062 & 0.019 & 0.019 & 0.003 & 0.023 & 0.003 & 0.947 & 0.953\tabularnewline
		$\alpha_{3}$ & 1 & 0.944 & 0.989 & $-$0.056 & $-$0.011 & 0.118 & 0.095 & 0.121 & 0.095 & 0.946 & 0.958 \tabularnewline
		$\rho_{12}$ & 0.4 & 0.484 & 0.375 & 0.084 & $-$0.025 & 0.032 & 0.035 & 0.038 & 0.036 & 0.947 & 0.957  \tabularnewline
		$\rho_{13}$ & 0.4 & 0.426 & 0.434 & 0.026 & 0.034 & 0.071 & 0.045 & 0.071 & 0.046 & 0.950 & 0.954\tabularnewline
		$\rho_{23}$ & 0.4& 0.452 & 0.437 & 0.052 & 0.037 & 0.153 & 0.124 & 0.156 & 0.126 & 0.948 & 0.949\tabularnewline
		\midrule
		\multicolumn{12}{c}{Setting III}\tabularnewline
		$\beta_{1}$ & 1 &1.044 & 1.005 & 0.044 & 0.005 & 0.008 & 0.009 & 0.010 & 0.009 & 0.948 & 0.952\tabularnewline
		$\beta_{2}$ & 0.8& 0.801 & 0.841 & 0.001 & 0.041 & 0.066 & 0.009 & 0.066 & 0.010 & 0.956 & 0.958 \tabularnewline
		$\beta_{3}$ & 0.4& 0.423 & 0.401 & 0.023 & 0.001 & 0.108 & 0.099 & 0.108 & 0.099 & 0.953 & 0.956  \tabularnewline
		$\alpha_{1}$ & 1  & 1.067 & 1.016 & 0.067 & 0.016 & 0.117 & 0.054 & 0.122 & 0.054 & 0.948 & 0.952 \tabularnewline
		$\alpha_{2}$ & 1& 1.006 & 1.022 & 0.006 & 0.022 & 0.032 & 0.028 & 0.032 & 0.029 & 0.950 & 0.954\tabularnewline
		$\alpha_{3}$ & 1 & 1.041 & 1.026 & 0.041 & 0.025 & 0.135 & 0.058 & 0.136 & 0.059 & 0.946 & 0.962\tabularnewline
		$\rho_{12}$ & 0.8 & 0.873 & 0.851 & 0.073 & 0.050 & 0.141 & 0.115 & 0.146 & 0.118 & 0.946 & 0.954\tabularnewline
		$\rho_{13}$ & 0.8& 0.887 & 0.803 & 0.087 & 0.003 & 0.093 & 0.018 & 0.100 & 0.018 & 0.940 & 0.958\tabularnewline
		$\rho_{23}$ & 0.8& 0.711 & 0.793 & $-$0.089 & $-$0.007 & 0.146 & 0.011 & 0.154 & 0.011 & 0.949 & 0.951\tabularnewline
		\bottomrule
	\end{tabular}}
\end{table}

\begin{table}
	\caption{\label{tab:Empirical-standard-errors-Gaussian_Gaussian}Empirical results
		from simulation studies examining the properties of the estimators
		of treatment effects, variances, and copula parameter for the GG model (i.e.,
		 Gaussian copula with Gaussian marginal distributions). }

	\centering{}%
	{\small
	\begin{tabular}{clrrrrrrrrrr}
	\toprule
		Param. & Value  & \multicolumn{2}{c}{Mean} & \multicolumn{2}{c}{Bias} & \multicolumn{2}{c}{Var} & \multicolumn{2}{c}{$\MSE$} & \multicolumn{2}{c}{CP}\tabularnewline
		\midrule
		\multicolumn{2}{c}{\# of subjects} & 200 & 400 & 200 & 400 & 200 & 400 & 200 & 400 & 200 & 400\tabularnewline
		\midrule
		\multicolumn{12}{c}{Setting I}\tabularnewline
		$\beta_{1}$ & 1& 0.958 & 0.964 & $-$0.042 & $-$0.036 & 0.057 & 0.015 & 0.059 & 0.016 & 0.956 & 0.957 \tabularnewline
		$\beta_{2}$ & 0.8 & 0.746 & 0.776 & $-$0.054 & $-$0.024 & 0.095 & 0.033 & 0.098 & 0.034 & 0.953 & 0.958\tabularnewline
		$\beta_{3}$ & 0.4 & 0.426 & 0.424 & 0.026 & 0.024 & 0.098 & 0.055 & 0.099 & 0.055 & 0.946 & 0.956\tabularnewline
		$\alpha_{1}$ & 1& 1.054 & 1.001 & 0.053 & 0.001 & 0.080 & 0.031 & 0.082 & 0.031 & 0.949 & 0.959 \tabularnewline
		$\alpha_{2}$ & 1& 0.991 & 0.956 & $-$0.009 & $-$0.044 & 0.018 & 0.007 & 0.019 & 0.008 & 0.947 & 0.950\tabularnewline
		$\alpha_{3}$ & 1& 1.006 & 1.029 & 0.006 & 0.029 & 0.041 & 0.039 & 0.041 & 0.040 & 0.952 & 0.955\tabularnewline
		$\rho_{12}$ & 0 & 0.141 & 0.051 & 0.141 & 0.051 & 0.077 & 0.049 & 0.097 & 0.052 & 0.933 & 0.955\tabularnewline
		$\rho_{13}$ & 0& 0.093 & 0.042 & 0.093 & 0.042 & 0.125 & 0.058 & 0.134 & 0.060 & 0.933 & 0.956\tabularnewline
		$\rho_{23}$ & 0 &0.077 & 0.056 & 0.077 & 0.056 & 0.086 & 0.088 & 0.092 & 0.091 & 0.940 & 0.944\tabularnewline
		\midrule
		\multicolumn{12}{c}{Setting II}\tabularnewline
		$\beta_{1}$ & 1 & 0.935 & 0.987 & $-$0.065 & $-$0.013 & 0.048 & 0.009 & 0.053 & 0.009 & 0.959 & 0.959\tabularnewline
		$\beta_{2}$ & 0.8& 0.786 & 0.841 & $-$0.014 & 0.041 & 0.115 & 0.113 & 0.115 & 0.114 & 0.949 & 0.960\tabularnewline
		$\beta_{3}$ & 0.4 & 0.455 & 0.449 & 0.055 & 0.049 & 0.069 & 0.011 & 0.072 & 0.013 & 0.952 & 0.965\tabularnewline
		$\alpha_{1}$ & 1& 1.002 & 1.025 & 0.002 & 0.025 & 0.041 & 0.012 & 0.041 & 0.013 & 0.950 & 0.955\tabularnewline
		$\alpha_{2}$ & 1 & 0.980 & 0.966 & $-$0.020 & $-$0.034 & 0.098 & 0.000 & 0.098 & 0.001 & 0.948 & 0.949\tabularnewline
		$\alpha_{3}$ & 1 & 0.964 & 1.006 & $-$0.036 & 0.005 & 0.116 & 0.098 & 0.117 & 0.098 & 0.945 & 0.953 \tabularnewline
		$\rho_{12}$ & 0.4 & 0.494 & 0.430 & 0.094 & 0.030 & 0.101 & 0.044 & 0.110 & 0.045 & 0.941 & 0.957 \tabularnewline
		$\rho_{13}$ & 0.4 & 0.489 & 0.437 & 0.089 & 0.037 & 0.072 & 0.018 & 0.080 & 0.020 & 0.947 & 0.949\tabularnewline
		$\rho_{23}$ & 0.4& 0.494 & 0.430 & 0.094 & 0.030 & 0.101 & 0.044 & 0.110 & 0.045 & 0.941 & 0.957\tabularnewline
		\midrule
		\multicolumn{12}{c}{Setting III}\tabularnewline
		$\beta_{1}$ & 1 & 0.975 & 1.013 & $-$0.025 & 0.013 & 0.149 & 0.069 & 0.149 & 0.069 & 0.959 & 0.952\tabularnewline
		$\beta_{2}$ & 0.8& 0.840 & 0.848 & 0.040 & 0.048 & 0.139 & 0.038 & 0.140 & 0.041 & 0.951 & 0.957  \tabularnewline
		$\beta_{3}$ & 0.4& 0.357 & 0.449 & $-$0.043 & 0.049 & 0.074 & 0.007 & 0.076 & 0.009 & 0.952 & 0.962 \tabularnewline
		$\alpha_{1}$ & 1  & 1.016 & 1.010 & 0.016 & 0.009 & 0.032 & 0.031 & 0.032 & 0.031 & 0.950 & 0.956\tabularnewline
		$\alpha_{2}$ & 1& 1.012 & 1.019 & 0.012 & 0.019 & 0.085 & 0.066 & 0.085 & 0.067 & 0.952 & 0.958\tabularnewline
		$\alpha_{3}$ & 1 & 0.950 & 1.020 & $-$0.050 & 0.020 & 0.028 & 0.017 & 0.031 & 0.018 & 0.946 & 0.954\tabularnewline
		$\rho_{12}$ & 0.8 & 0.848 & 0.834 & 0.048 & 0.034 & 0.120 & 0.039 & 0.123 & 0.040 & 0.944 & 0.951\tabularnewline
		$\rho_{13}$ & 0.8 & 0.825 & 0.801 & 0.025 & 0.001 & 0.131 & 0.100 & 0.132 & 0.100 & 0.942 & 0.957\tabularnewline
		$\rho_{23}$ & 0.8 & 0.872 & 0.792 & 0.072 & $-$0.008 & 0.052 & 0.052 & 0.057 & 0.052 & 0.942 & 0.948\tabularnewline
		\bottomrule
	\end{tabular}}
\end{table}

\section{Application to a Skin Cancer Dataset\label{sec:Application-to-Real_Co_ch}}
The Nutritional Prevention of Cancer (NPC) study was a randomized,
double-blinded, placebo-controlled clinical trial. The goal of the NPC study was to evaluate
the efficacy of selenium  in preventing the recurrence of nonmelanoma skin cancers. There were 1312 individuals from the Eastern United States in the study. Full details of the study design and major results are available in \shortciteN{duffield2002baseline}, and \shortciteN{duffield2003selenium}.
Participants were recruited from seven dermatology practices located in cities in low-selenium areas.  Participants had a history of two or more BCCs or one SCC of the skin, with one of these developed within the year before the study enrollment.

Subjects were randomized in a double-blinded fashion to either 200 $\mu$g/day of selenium in 0.5-gram high selenium baker's yeast or a  placebo. Recruitment was between 1983 and 1991. Only subjects with valid baseline plasma selenium values collected on the day of randomization, plus or minus four days, were included. Sixty-two participants whose initial blood draws were not drawn within four days of the randomization date, and an additional 58 subjects who were found to have incomplete follow-up data are excluded from the analysis. As a result, our study was based on data from the 1192 participants with valid baseline selenium values (i.e., there are 606 participants in the selenium group and 586 participants in the placebo group).

\begin{figure}
	\includegraphics[width=1\textwidth,height=0.2\textheight]{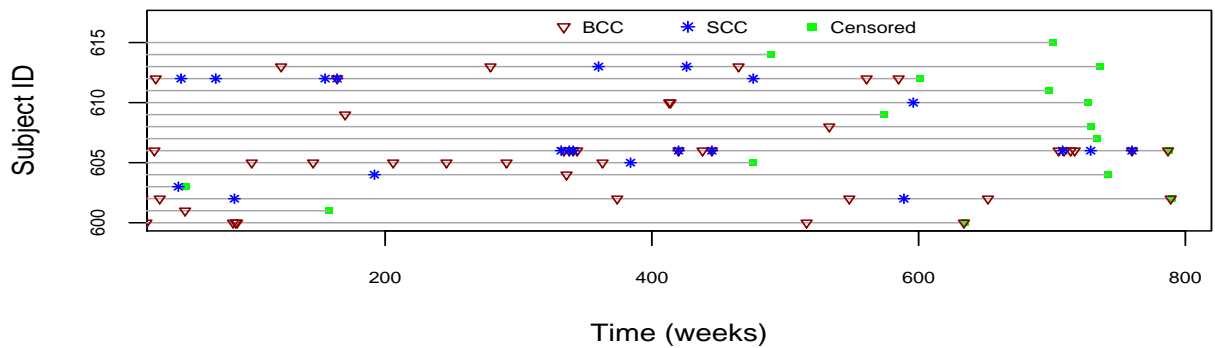}
	\caption{\label{fig:The-profile-plot_Copula_ch}The event plot for subjects
		600 to 615.}
\end{figure}

The follow up extended from the date of randomization through to February 1, 1996. The mean follow-up time was 7.4 years.
The total number of BCCs and SCCs are 1582 and 351 for the placebo group, and 1871 and 408 for the selenium group, respectively. The average number of BCCs per subject by the treatment group (selenium vs placebo) are 2.699 and 3.087.  The average number of SCCs per subject by the treatment group are 0.599 and 0.673. There were 25.9\% and 66.7\% of the participants who did not develop new BCC or SCC, respectively, during the study. The length of the follow-up and the recurrent times of the BCC and SCC for a sample of 15 patients are depicted in Figure~\ref{fig:The-profile-plot_Copula_ch}.

This study intends to estimate the effects of selenium supplementation
on the recurrence of BCC and SCC cancer types. Because
the correlation between two processes is of key interest, it is necessary to use a multi-type recurrent event model. Another important point of interest is to study whether the subjects
who are at higher risk of BCC tend to have a higher or lower risk
of SCC. This approach allows us to assess the association
between the recurrence of  BCCs and SCCs, and the dependence of
within-subject events. The models presented in Section \ref{sec:The-model} were applied to this skin cancer dataset. There is one treatment covariate with two levels at  $x_i= 1$  and  $x_i = 0$ (i.e., selenium group vs placebo group, respectively).  Regression coefficients for the selenium effects on BCC and SCC are $\beta_{1}$ and $\beta_{2}$ , respectively.  Variance components of frailties $(\alpha_{1}, \alpha_{2}$), and the copula parameter ($\alpha$) are estimated.

We consider the four copula models as listed in Table~\ref{tab:model.label}. Table \ref{tab:Parameter-estimates-for_cla_Gam_cop-Ch} gives the model parameter estimates. Here we discuss the results from each model.

\begin{inparaitem}
\item Based on the Clayton copula and gamma frailty model (Model Cg),
it is suggested that the intensity of BCC increases in the selenium group compared to the placebo group
with a relative risk (RR) of 1.088. The selenium's group is associated with
a higher risk of SCC recurrence (RR = 1.119). The
estimated variances ($\alpha_{1}$ and $\alpha_{2})$ are 1.045
and 2.686, respectively. This implies that the
within-subject correlation of SCC recurrent times
is stronger than the recurrent times of  BCC.
However, the between-subject variation of SCC times is higher than the between-subject variation of BCC times. A moderate correlation between
the event times for BCC and SCC was observed (Kendall's tau equals 0.3). The positive estimates of the correlation
between the two processes indicate that a higher event rate of BCC tends to occur with a higher event rate of SCC.

\item Based on the Clayton copula and Gaussian
random effects model (Model CG), it
shows that selenium supplementation was associated with an increase in recurrence
of both BCC and  SCC with RR as 1.093 and 1.131,
respectively. The estimated variances and Kendall's tau are
1.001, 1.978, and 0.345, respectively.

\item Based on the Gaussian copula and gamma frailties model (Model
Gg), the rate of the BCC is higher
in the selenium group than the placebo group (RR = 1.090), and the incidence
rate is higher (RR = 1.124) in the selenium group compared to the placebo group for SCC type. There is a significant
heterogeneity for both the BCC and SCC event types with $\alpha_{1}$ = 1.057
and $\alpha_{2}$ = 2.798. The correlation coefficient between the two event
types is estimated at 0.296.

\item Based on the Gaussian copula and Gaussian random
effects model (Model GG). The rate of recurrence of BCC (RR = 1.111) and SCC (RR = 1.163) is
increased with the selenium group relative to the placebo group.
The positive values of $\alpha_{1}$ = 1.027, $\alpha_{2}$ = 2.044, and
$\alpha$ = 0.315 indicate dependence within the subject event times
and between the two cancer types.
\end{inparaitem}

\begin{table}
	\caption{\label{tab:Parameter-estimates-for_cla_Gam_cop-Ch}Parameter estimates
		for the multi-type recurrent events models. }
	
	\centering
	\begin{tabular}{c|cccc|cccc}\hline\toprule
		\multirow{2}{*}{Parameter} & \multicolumn{4}{c|}{Model Cg} & \multicolumn{4}{c}{Model CG} \tabularnewline
		& $\mbox{EST.}$ & $\mbox{S.E.}$ & $\mbox{RR}$ & $\mbox{p-value}$ & $\mbox{EST.}$ & $\mbox{S.E.}$ & $\mbox{RR}$ & $\mbox{p-value}$   \tabularnewline
		\hline
		$\beta_{1}$   & 0.085 & 0.072 & 1.088 & 0.239 & 0.091 & 0.076 & 1.093 & 0.242 \tabularnewline
		$\beta_{2}$   & 0.113 & 0.103 & 1.119 & 0.273 & 0.124 & 0.112 & 1.131 & 0.269 \tabularnewline
		$\alpha_{1}$  & 1.045 & 0.076 &  & $<$0.001 & 1.001 & 0.069  &  & $<$0.001  \tabularnewline
		$\alpha_{2}$  & 2.686  & 0.153 &  & $<$0.001 & 1.978 & 0.135 &  & $<$0.001  \tabularnewline
		$\alpha$      & 0.856 & 0.098 &  & $<$0.001 & 1.076 & 0.096 &  & $<$0.001   \tabularnewline
		Kendall's tau & 0.300 &  &  &  & 0.345 &  &  &    \tabularnewline\hline
		& \multicolumn{4}{c|}{Model Gg} & \multicolumn{4}{c}{Model GG} \tabularnewline
		& $\mbox{EST.}$ & $\mbox{S.E.}$ & $\mbox{RR}$ & $\mbox{p-value}$ & $\mbox{EST.}$ & $\mbox{S.E.}$ & $\mbox{RR}$ & $\mbox{p-value}$\tabularnewline\hline
		$\beta_{1}$   & 0.084 & 0.070 & 1.090 & 0.219 & 0.088 & 0.072 & 1.111 & 0.221  \tabularnewline
		$\beta_{2}$   & 0.117  & 0.106 & 1.124 & 0.120 & 0.122 & 0.128 & 1.163 & 0.340 \tabularnewline
		$\alpha_{1}$  & 1.057 & 0.074 &  & $<$0.001 & 1.027 & 0.072 &  & $<$0.001\tabularnewline
		$\alpha_{2}$  & 2.798 & 0.125 &  & $<$0.001 & 2.044 & 0.089 &  & $<$0.001\tabularnewline
		$\alpha$      & 0.296 & 0.089 &  & $<$0.001 & 0.315 & 0.080 &  & $<$0.001\tabularnewline
		\bottomrule
	\end{tabular}
\end{table}

In summary, results from all models in Tables \ref{tab:Parameter-estimates-for_cla_Gam_cop-Ch} show that the variance of the random effects and frailties for SCC event times within-subject  is, however, much larger than that for BCC. The correlation coefficient $\alpha$
and Kendall's tau are also significantly different from zero,
indicating a moderate association between the risk of BCC and SCC
recurrences. Based on the calculated sum of squared deviance residuals (e.g., \citeNP{therneau2000modeling}) in Table~\ref{tab:Deviance-residuals-for_Cop_ch}, Model Gg fits the skin cancer data better than other models (Cg, CG, GG). The Gaussian copula describes symmetric dependence between the subject frailties of BCC and SCC cancer types with less tail dependence. Gamma marginals confirm a heavy lower tail and relatively high density for subjects with large frailties of the two cancer types. It is clear that assuming different models of frailties/random effects tends not to have a considerable effect on the inference of model parameters in Table \ref{tab:Parameter-estimates-for_cla_Gam_cop-Ch}.

\begin{figure}
	\includegraphics[width=0.5\textwidth,height=0.3\textheight]{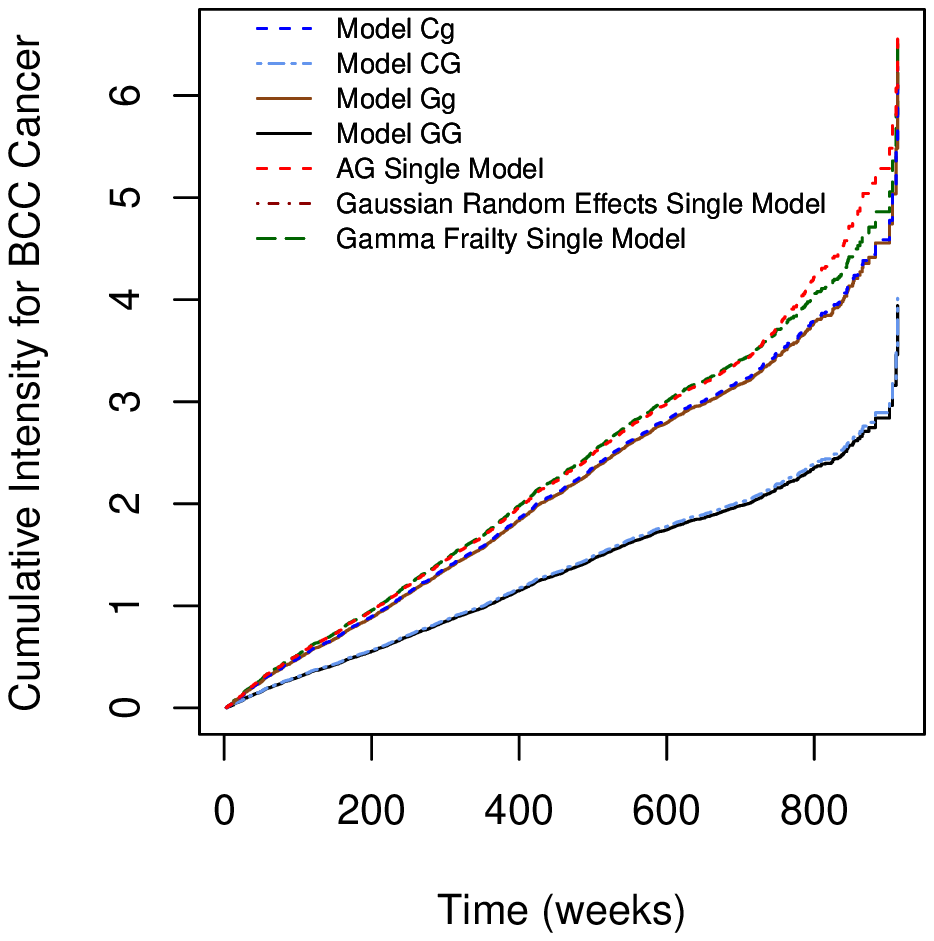}
\includegraphics[width=0.5\textwidth,height=0.3\textheight]{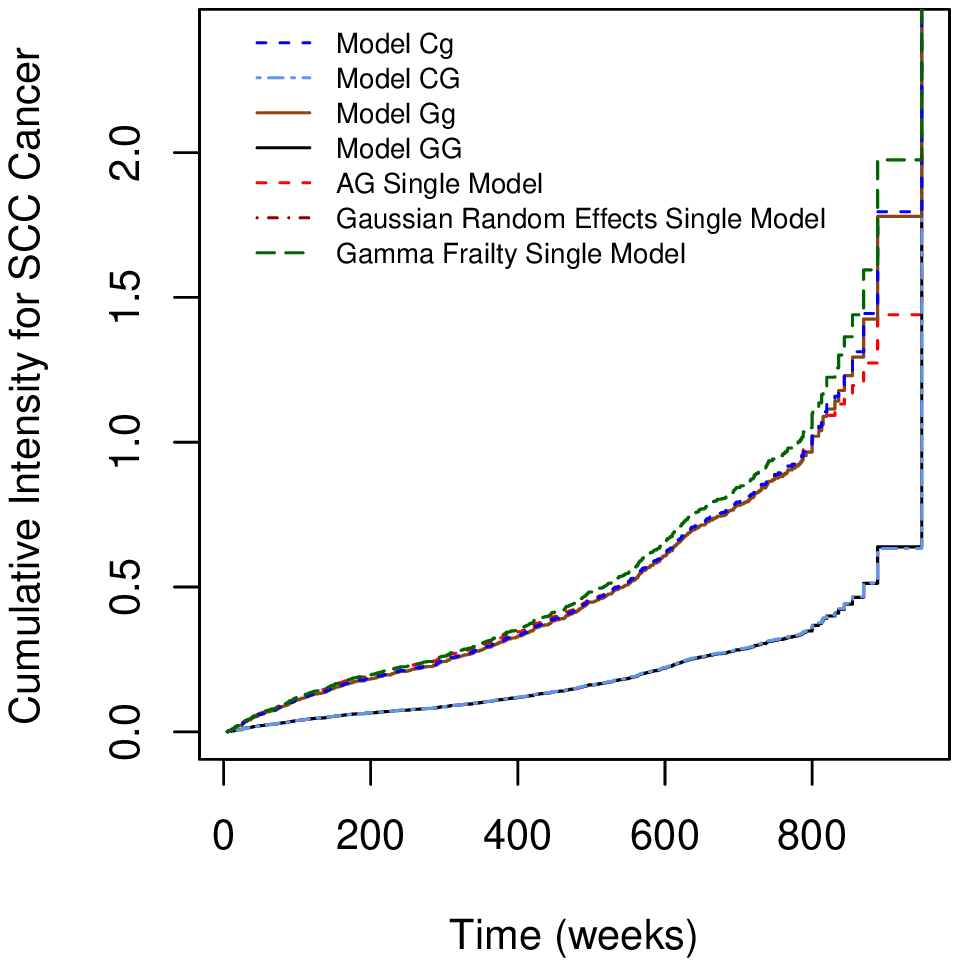}
	\caption{\label{fig:Cumulative-intensity-functions_type1_cop_ch} Cumulative
		intensity functions for the BCC and SCC tumor types.}
\end{figure}

\begin{table}
	\caption{\label{tab:Deviance-residuals-for_Cop_ch}Deviance residuals for the fitted
		multi-type recurrent events models.}

	\centering{}%
	\begin{tabular}{ccccc}
		\toprule
		Type & Model Cg & Model CG & Model Gg & Model GG\tabularnewline
		\hline
		BCC & \textcolor{black}{355} & 400 & \textcolor{black}{349} & 394\tabularnewline
		SCC & \textcolor{black}{307} & 326 & \textcolor{black}{300} & 333\tabularnewline
		\hline
		Total  & 662  & 726 & 649 & 727\tabularnewline
		\bottomrule
	\end{tabular}
\end{table}

The Nelson-Aalen estimates of the cumulative baseline intensity functions
for both BCC and SCC obtained from a separate analysis using Cox models and
multi-type recurrent event models are shown in Figure \ref{fig:Cumulative-intensity-functions_type1_cop_ch}.
It shows  that the cumulative baseline intensity for BCC is higher
than the cumulative function for SCC. In addition, the estimated  baseline cumulative intensity functions for gamma frailty are higher than that for Gaussian random effects for both
Clayton and Gaussian copulas. However, the cumulative intensity functions in Figure~\ref{fig:Cumulative-intensity-functions_type1_cop_ch} are sensitive to the selection of the model of frailties/random effects (e.g., \citeNP{cook2010copula}).

\section{Conclusions and Areas for Future Work\label{sec:Conclusions_Co_ch}}
In this paper, we propose copula-frailty models for multi-type recurrent event data. We use copulas to characterize complicated dependence structures of multivariate frailties, which provides flexibility in modeling recurrent events with multi-types. Maximum likelihood estimates of treatment effects, variance components, and nonparametric cumulative baseline intensity functions are obtained. We implement the MCEM algorithms in the E-step and numerical maximization methods in the M-step. We apply the developed method to analyze the skin cancer data.

Although we only consider two copula functions (i.e., Gaussian and Clayton) and two marginal distributions (i.e., the Gaussian and gamma distributions) in this paper, it is relatively straightforward to extend the same estimation method to other copula functions and marginal distributions. Expanding the frailty model in \eqref{eq:model}  to accommodate random slopes needs further research, and it is interesting to investigate this in future research.

Automated MCEM algorithms (e.g., \citeNP{booth1999maximizing}) can be used for the current model to study the efficacy of different routines to minimize the computing expenses. We assume time-constant coefficients in the proposed models in \eqref{eq:model} and \eqref{eq:log normal model}. It is interesting to evaluate the long-term effects of risk factors that could vary with time.  We are currently extending the developed model and estimation procedure to accommodate for potential time varying-covariate effects and for nonparametric covariates as well.


\section*{Acknowledgments}
The authors thank the editor, associate editor, and referee, for their valuable comments that helped in improving the paper significantly. The authors acknowledge Advanced Research Computing at Virginia Tech for providing computational resources. The work by Hong was partially supported by National Science Foundation Grant CMMI-1904165 to Virginia Tech.


\appendix
\section{Metropolis-Hastings Algorithm}\label{sec:B.1Metropolis-Hasting-Algorithm}

In this section, we briefly describe the Metropolis-Hastings algorithm.
Our approach involves generating random samples $\wvec_{i}$
from the exact conditional distribution of the frailty terms (given
the data) by MCMC sampling. From \eqref{eq:marginal frai}, the conditional
distribution of $\wvec_{i}|\mathbf{D}_{i}$ can be expressed
as
\[
g_{\wvec_{i}|\mathbf{D}_{i}}(\wvec_{i}|\xivec)\propto L_{i}(\xivec,\wvec_{i}).
\]
There are several approaches to select proposed functions, resulting
in specific types of Metropolis-Hastings algorithms. We use a random-walk Metropolis-Hastings
algorithm. For the density distribution of the candidates, we use copula
density functions that match copula densities for the frailty terms
$\wvec_{i}$ to propose candidate samples for $\wvec_{i}^{\ast}$.
For example, Gaussian and Clayton copulas and Gaussian marginal distributions
with some proposed copula parameter values
are used. The Metropolis-Hastings algorithm is described in \textbf{Algorithm~\ref{alg:A-Metropolis-Hastings-algorithm-1}}.

\begin{algorithm}
	\caption{\label{alg:A-Metropolis-Hastings-algorithm-1}A Metropolis-Hastings
		algorithm for the E-step in the MCEM algorithm.}
	\begin{enumerate}
		\item Initialize $\wvec^{(0)}=(\wvec_{1}^{(0)\prime},\ldots,\wvec_n^{(0)\prime})'$.
		At iteration $q,\,(q=1,\ldots, n_q)$,
		\item Sample $\wvec_{i}^{\ast}$ , for the $i^{th}$ component
		of $\wvec$ from a proposed distribution $\wvec_{i}^{\ast}\sim g(\wvec_{i}^{\ast}|\wvec_{i}^{(q-1)})$.
		\item Compute an acceptance ratio (probability) \\
		\[r=\frac{g_{\wvec_{i}|\mathbf{D}_{i}}(\wvec_{i}^{\ast}|\xivec)g(\wvec_{i}^{(q-1)}|\wvec_{i}^{\ast})}
		{g_{\wvec_{i}|\mathbf{D}_{i}}(\wvec_{i}^{(q-1)}|\xivec)g(\wvec_{i}^{\ast}|\wvec_{i}^{(q-1)})}.
		\]
		\item Sample $u\,\sim$ uniform $(0,1).$ Set $\wvec_{i}^{(q)}$
		to $\wvec_{i}^{\ast}$ if $u<r$ and to $\wvec_{i}^{(q-1)}$
		if $u>r$.
		\item Repeat steps $2$ to $4$ for generating $n_q$ random samples $\wvec^{(1)},\ldots,\wvec^{(n_q)}$.
	\end{enumerate}
\end{algorithm}

\section{Expected Log-likelihood of Gaussian and Clayton Copula with Gamma Marginals}\label{sec:B.2Expected-Log-likelihood-of}
In this section, we derive the expected log-likelihood $Q_{2}(\alphavec)$
for Gaussian and Clayton copula with gamma marginal distributions.
For instance, when frailty terms are Gaussian copula, $Q_{3}(\alphavec_c)$
is
\begin{eqnarray*}
	Q_{3}(\alphavec_c) & = & -\frac{1}{2}\log(| \Rvec_{m}|)-\frac{1}{2}\mbox{E}\Big[\,\boldsymbol{q}_{i}'(\Rvec_{m}^{-1}-\Ivec_{m})\boldsymbol{q}_{i}\Big]\\
	& = & -\frac{1}{2}\log(| \Rvec_{m}|)-\frac{1}{2}\mbox{tr}\Big[(\Rvec_{m}^{-1}-\Ivec_{m})\mbox{E}(\boldsymbol{q}_{i}'\boldsymbol{q}_{i})\Big].
\end{eqnarray*}
Note that $\mbox{E}(\cdot)$ is the expectation
of the frailty conditional distribution. In such case when we consider
a Clayton copula, $Q_{3}(\alphavec_c)$ can be expressed as
\begin{multline*}
Q_{3}(\alphavec_c)=\sum_{i=1}^n\left\{ (-1/\alpha-m)\mbox{E}\Big[\log(\sum_{j=1}^{m}u_{ij}^{-\alpha}-m+1)\Big]+\sum_{j=1}^{m}\mbox{E}\Big[\log(u_{ij}^{-\alpha-1})\Big]\right\} \\
+n\,\log\left\{ (-\alpha)^{m}\prod_{j=0}^{m-1}(-\frac{1}{\alpha}-j)\right\} .
\end{multline*}
For example, a bivariate Clayton copula, $Q_{3}(\alphavec_c)$
is
\begin{multline*}
g(w_{i1},\,w_{i2})=\sum_{i=1}^n\left\{ (-1/\alpha-2)\mbox{E}[\log(u_{i1}^{-\alpha}+u_{i2}^{-\alpha}-1)]+(-\alpha-1)\mbox{E}[\log(u_{i1})+\log(u_{i2})]\right\} \\
+n[\log(1+\alpha)].
\end{multline*}
For marginal distributions, when $w_{ij}$ are gamma$(1/\alpha_{j},\,\alpha_{j})$
distributed with mean 1 and variance $\alpha_{j}$,
$Q_{4}(\alpha_{j})$ can be in the form
\[Q_{4}(\alpha_{j})=\sum_{i=1}^n\left\{ \left(\frac{1}{\alpha_{j}}-1\right)\mbox{E}\big[\log(w_{ij})\big]-\frac{1}{\alpha_{j}}\mbox{E\ensuremath{(w_{ij})}}\right\} -n\left\{\log\Big[\Gamma\left(\frac{1}{\alpha_{j}}\right)\Big]+\frac{1}{\alpha_{j}}\log(\alpha_{j})\right\}.
\]



\begin{thebibliography}{}

\bibitem[\protect\citeauthoryear{Andersen and Gill}{Andersen and
  Gill}{1982}]{andersen1982cox}
Andersen, P.~K. and R.~D. Gill (1982).
\newblock Cox's regression model for counting processes: a large sample study.
\newblock {\em The {A}nnals of {S}tatistics\/}~{\em 10}, 1100--1120.

\bibitem[\protect\citeauthoryear{Bedair, Hong, Li, and Al-Khalidi}{Bedair
  et~al.}{2016}]{bedair2016multivariate}
Bedair, K., Y.~Hong, J.~Li, and H.~R. Al-Khalidi (2016).
\newblock Multivariate frailty models for multi-type recurrent event data and
  its application to cancer prevention trial.
\newblock {\em Computational Statistics \& Data Analysis\/}~{\em 101},
  161--173.

\bibitem[\protect\citeauthoryear{Booth and Hobert}{Booth and
  Hobert}{1999}]{booth1999maximizing}
Booth, J.~G. and J.~P. Hobert (1999).
\newblock Maximizing generalized linear mixed model likelihoods with an
  automated {M}onte {C}arlo {EM} algorithm.
\newblock {\em {J}ournal of the Royal Statistical Society: Series B
  (Statistical Methodology)\/}~{\em 61}, 265--285.

\bibitem[\protect\citeauthoryear{Chatterjee and Roy}{Chatterjee and
  Roy}{2018}]{ChatterjeeRoy2018}
Chatterjee, M. and S.~S. Roy (2018).
\newblock A copula-based approach for estimating the survival functions of two
  alternating recurrent events.
\newblock {\em Journal of Statistical Computation and Simulation\/}~{\em 88},
  3098--3115.

\bibitem[\protect\citeauthoryear{Cook, Lawless, and Lee}{Cook
  et~al.}{2010}]{cook2010copula}
Cook, R.~J., J.~F. Lawless, and K.-A. Lee (2010).
\newblock A copula-based mixed {P}oisson model for bivariate recurrent events
  under event-dependent censoring.
\newblock {\em Statistics in {M}edicine\/}~{\em 29}, 694--707.

\bibitem[\protect\citeauthoryear{Duchateau and Janssen}{Duchateau and
  Janssen}{2008}]{duchateau_frailty_2008}
Duchateau, L. and P.~Janssen (2008).
\newblock {\em The Frailty Model}.
\newblock New York: Springer.

\bibitem[\protect\citeauthoryear{Duffield-Lillico, Reid, Turnbull, Combs,
  Slate, Fischbach, Marshall, and Clark}{Duffield-Lillico
  et~al.}{2002}]{duffield2002baseline}
Duffield-Lillico, A.~J., M.~E. Reid, B.~W. Turnbull, G.~F. Combs, E.~H. Slate,
  L.~A. Fischbach, J.~R. Marshall, and L.~C. Clark (2002).
\newblock Baseline characteristics and the effect of selenium supplementation
  on cancer incidence in a randomized clinical trial: a summary report of the
  nutritional prevention of cancer trial.
\newblock {\em Cancer Epidemiology and Prevention Biomarkers\/}~{\em 11},
  630--639.

\bibitem[\protect\citeauthoryear{Duffield-Lillico, Slate, Reid, Turnbull,
  Wilkins, Combs, Park, Gross, Graham, Stratton, et~al.}{Duffield-Lillico
  et~al.}{2003}]{duffield2003selenium}
Duffield-Lillico, A.~J., E.~H. Slate, M.~E. Reid, B.~W. Turnbull, P.~A.
  Wilkins, G.~F. Combs, H.~K. Park, E.~G. Gross, G.~F. Graham, M.~S. Stratton,
  et~al. (2003).
\newblock Selenium supplementation and secondary prevention of nonmelanoma skin
  cancer in a randomized trial.
\newblock {\em {J}ournal of the National Cancer Institute\/}~{\em 95},
  1477--1481.

\bibitem[\protect\citeauthoryear{Joe}{Joe}{2005}]{Joe2005}
Joe, H. (2005).
\newblock Asymptotic efficiency of the two-stage estimation method for
  copula-based models.
\newblock {\em Journal of Multivariate Analysis\/}~{\em 94}, 401--419.

\bibitem[\protect\citeauthoryear{Joe and Xu}{Joe and Xu}{1996}]{JoeXu1996}
Joe, H. and J.~J. Xu (1996).
\newblock The estimation method of inference functions for margins for
  multivariate models.
\newblock Technical report, University of British Columbia,
  doi:10.14288/1.0225985.

\bibitem[\protect\citeauthoryear{Lee and Cook}{Lee and
  Cook}{2019}]{LeeCook2019}
Lee, J. and R.~J. Cook (2019).
\newblock Dependence modeling for multi-type recurrent events via copulas.
\newblock {\em Statistics in Medicine\/}~{\em 38}, 4066--4082.

\bibitem[\protect\citeauthoryear{Li, Guo, and Kim}{Li
  et~al.}{2020}]{LiGuoKim2020}
Li, Q., F.~Guo, and I.~Kim (2020).
\newblock A non-parametric {Bayesian} change-point method for recurrent events.
\newblock {\em Journal of Statistical Computation and Simulation\/}~{\em 90},
  2929--2948.

\bibitem[\protect\citeauthoryear{Lin, Luo, Chen, and Davis}{Lin
  et~al.}{2017}]{lin2017bayesian}
Lin, L.-A., S.~Luo, B.~E. Chen, and B.~R. Davis (2017).
\newblock Bayesian analysis of multi-type recurrent events and dependent
  termination with nonparametric covariate functions.
\newblock {\em Statistical Methods in Medical Research\/}~{\em 26}, 2869--2884.

\bibitem[\protect\citeauthoryear{Liu and Huang}{Liu and
  Huang}{2008}]{liu2008use}
Liu, L. and X.~Huang (2008).
\newblock The use of {G}aussian quadrature for estimation in frailty
  proportional hazards models.
\newblock {\em Statistics in Medicine\/}~{\em 27}, 2665--2683.

\bibitem[\protect\citeauthoryear{Louis}{Louis}{1982}]{louis1982finding}
Louis, T.~A. (1982).
\newblock Finding the observed information matrix when using the {EM}
  algorithm.
\newblock {\em {J}ournal of the Royal Statistical Society. Series B
  (Methodological)\/}~{\em 22}, 226--233.

\bibitem[\protect\citeauthoryear{Mazroui, Mauguen, Mathoulin-P{\'e}lissier,
  MacGrogan, Brouste, and Rondeau}{Mazroui et~al.}{2016}]{mazroui2015time}
Mazroui, Y., A.~Mauguen, S.~Mathoulin-P{\'e}lissier, G.~MacGrogan, V.~Brouste,
  and V.~Rondeau (2016).
\newblock Time-varying coefficients in a multivariate frailty model:
  Application to breast cancer recurrences of several types and death.
\newblock {\em Lifetime Data Analysis\/}~{\em 22}, 1--25.

\bibitem[\protect\citeauthoryear{Nelsen}{Nelsen}{1999}]{nelsen1999introduction}
Nelsen, R.~B. (1999).
\newblock {\em An{ I}ntroduction to{ C}opulas}.
\newblock New York: Springer.

\bibitem[\protect\citeauthoryear{Parner et~al.}{Parner
  et~al.}{1998}]{parner1998asymptotic}
Parner, E. et~al. (1998).
\newblock Asymptotic theory for the correlated gamma-frailty model.
\newblock {\em The Annals of Statistics\/}~{\em 26}, 183--214.

\bibitem[\protect\citeauthoryear{Prentice, Williams, and Peterson}{Prentice
  et~al.}{1981}]{prentice1981regression}
Prentice, R.~L., B.~J. Williams, and A.~V. Peterson (1981).
\newblock On the regression analysis of multivariate failure time data.
\newblock {\em Biometrika\/}~{\em 68}, 373--379.

\bibitem[\protect\citeauthoryear{Rondeau, Mathoulin-Pelissier, Jacqmin-Gadda,
  Brouste, and Soubeyran}{Rondeau et~al.}{2007}]{rondeau2007joint}
Rondeau, V., S.~Mathoulin-Pelissier, H.~Jacqmin-Gadda, V.~Brouste, and
  P.~Soubeyran (2007).
\newblock Joint frailty models for recurring events and death using maximum
  penalized likelihood estimation: application on cancer events.
\newblock {\em Biostatistics\/}~{\em 8}, 708--721.

\bibitem[\protect\citeauthoryear{Shih and Louis}{Shih and
  Louis}{1995}]{shih1995inferences}
Shih, J.~H. and T.~A. Louis (1995).
\newblock Inferences on the association parameter in copula models for
  bivariate survival data.
\newblock {\em Biometrics\/}, 1384--1399.

\bibitem[\protect\citeauthoryear{Tawiah, McLachlan, and Ng}{Tawiah
  et~al.}{2020}]{TawiahMcLachlanNg2020}
Tawiah, R., G.~J. McLachlan, and S.~K. Ng (2020).
\newblock A bivariate joint frailty model with mixture framework for survival
  analysis of recurrent events with dependent censoring and cure fraction.
\newblock {\em Biometrics\/}~{\em 76}, 753--766.

\bibitem[\protect\citeauthoryear{Therneau and Grambsch}{Therneau and
  Grambsch}{2000}]{therneau2000modeling}
Therneau, T.~M. and P.~M. Grambsch (2000).
\newblock {\em Modeling Survival Data: Extending the Cox Model}.
\newblock New York: Springer.

\bibitem[\protect\citeauthoryear{Wei, Lin, and Weissfeld}{Wei
  et~al.}{1989}]{wei1989regression}
Wei, L.-J., D.~Y. Lin, and L.~Weissfeld (1989).
\newblock Regression analysis of multivariate incomplete failure time data by
  modeling marginal distributions.
\newblock {\em {J}ournal of the American Statistical Association\/}~{\em 84},
  1065--1073.

\bibitem[\protect\citeauthoryear{Zeng, Ibrahim, Chen, Hu, and Jia}{Zeng
  et~al.}{2014}]{zeng2014multivariate}
Zeng, D., J.~G. Ibrahim, M.-H. Chen, K.~Hu, and C.~Jia (2014).
\newblock Multivariate recurrent events in the presence of multivariate
  informative censoring with applications to bleeding and transfusion events in
  myelodysplastic syndrome.
\newblock {\em Journal of Biopharmaceutical Statistics\/}~{\em 24}, 429--442.

\end{thebibliography}

\end{document}